\documentclass[twocolumn,showpacs,preprintnumbers,amsmath,amssymb]{revtex4}
\usepackage{graphicx}

\begin{document}

\title{Constraints on exotic spin-dependent interactions between electrons from helium fine-structure spectroscopy}

\author{Filip Ficek$^1$}
 \email{filip.ficek@student.uj.edu.pl}
\author{Derek F. Jackson Kimball$^2$}
\author{Mikhail G. Kozlov$^{3,4}$}
\author{Nathan Leefer$^5$}
\author{Szymon Pustelny$^{1}$}
\author{Dmitry Budker$^{5,6,7}$}
\affiliation{
$^1$ Institute of Physics, Jagiellonian University, \L{}ojasiewicza 11, 30-348 Krak\'{o}w, Poland\\
$^2$ Department of Physics, California State University - East Bay, Hayward, California 94542-3084, USA\\
$^3$ Petersburg Nuclear Physics Institute, Gatchina 188300, Russia\\
$^4$ St. Petersburg Electrotechnical University “LETI”, Prof. Popov Str. 5, 197376 St. Petersburg\\
$^5$ Helmholtz Institute Mainz, Johannes Gutenberg University, 55099 Mainz, Germany\\
$^6$ Department of Physics, University of California at Berkeley, Berkeley, California 94720-7300, USA\\
$^7$ Nuclear Science Division, Lawrence Berkeley National Laboratory, Berkeley, California 94720, USA
}

\date{\today}

\begin{abstract}
Agreement between theoretical calculations of atomic structure and spectroscopic measurements is used to constrain possible contribution of exotic spin-dependent interactions between electrons to the energy differences between states in helium-4. In particular, constraints on dipole-dipole interactions associated with the exchange of pseudoscalar bosons (such as axions or axion-like particles, ALPs) with masses $10^{-2}~{\rm eV} \lesssim m \lesssim 10^{4}~{\rm eV}$ are improved by a factor of $\sim100$. The first atomic-scale constraints on several exotic velocity-dependent dipole-dipole interactions are established as well.
%We analyze constraints on exotic spin-dependent interactions using spectroscopic measurements of $^4$He and theoretical calculations of the electronic energy-level structure. By comparing experimental and theoretical (quantum-electrodynamics based) results, we determine the maximum possible energy perturbations caused by exotic electron-electron interactions and hence derive constraints on anomalous electron-electron coupling in the system. Since this approach uses $^4$He, which is a more thoroughly studied system than previously used positronium, it enables more stringent constraints on such interactions from previously reported limits.
\end{abstract}


\maketitle

%\section{Introduction}
Heretofore undiscovered spin-dependent interactions \cite{Moo84,Dob06} naturally arise in theories predicting new bosons such as axions \cite{Wei78,Wil78,Din81,Shi80,Kim79,Zhi80}, familons \cite{Wil82,Gel83}, majorons \cite{Gel81,Chi81}, arions \cite{Ans82}, new spin-0 or spin-1 gravitons \cite{Sch79,Nev80,Nev82,Car94}, Kaluza-Klein zero modes in string theory \cite{Svr06,Arv10}, paraphotons \cite{Dob05}, and new $Z'$ bosons \cite{App03}. Such new bosons are connected to possible explanations of the nature of dark matter \cite{Ber05}, dark energy \cite{Fri03,Fla09}, the strong-CP problem \cite{Moo84}, and the hierarchy problem \cite{Gra15}.

The most commonly employed framework for the purpose of comparing different experimental searches for exotic spin-dependent interactions is that introduced in Ref.~\cite{Moo84} to describe long-range spin-dependent potentials associated with the axion and extended in Ref.~\cite{Dob06} to encompass long-range potentials associated with any generic spin-0 or spin-1 boson. The spin-dependent potentials enumerated in Ref.~\cite{Dob06} are characterized by dimensionless coupling constants that specify the strength of the interaction between various particles and a characteristic range $\lambda$ for the interaction associated with the reduced Compton wavelength of the new boson of mass $m_0$, $\lambda = \hbar/(m_0 c)$, where $\hbar$ is the reduced Planck constant and $c$ is the speed of light. Depending on the nature of the new interaction, different particles will have different coupling constants. In the present work, we study dipole-dipole interactions between electrons at the atomic scale through investigation of the electronic structure of helium-4.

Laboratory searches for exotic spin-dependent interactions mediated by new bosons are sensitive and broadly inclusive probes for global symmetries broken at high energy scales \cite{Moo84,Dob06}. For example, the fundamental properties of axions and the axion-like-particles (ALPs) mentioned above \cite{Moo84,Dob06,Wei78,Wil78,Din81,Shi80,Kim79,Zhi80,Wil82,Gel83,Gel81,Chi81,Ans82,Sch79,Nev80,Nev82,Car94,Svr06,Arv10} are characterized by a symmetry breaking scale $f_a$ and an interaction scale $\Lambda$. These scales determine, for example, the mass of the ALP
\begin{align}
m_0 c^2 = \frac{\Lambda^2}{f_a}~,
\end{align}
and the interaction of an ALP with a Standard Model fermion $X$ is proportional to $m_X c^2/f_a$ where $m_X$ is the fermion mass. In particular, this work improves laboratory constraints on exotic spin-spin forces between electrons mediated by bosons in the mass range between $10^{-2}$~eV and $10^4$~eV by two orders of magnitude. Our research is complementary to experiments searching for an axion/ALP coupling to photons, such as the Axion Dark Matter eXperiment (ADMX) \cite{Asz10}, the CERN Axion Solar Telescope (CAST) \cite{Zio05}, and light-shining-through-wall experiments such as the Any Light Particle Search (ALPS) \cite{Ehr10}, since He spectroscopy probes the electron-ALP interaction as opposed to the photon-ALP interaction and is sensitive to a mass range beyond that probed by experiments such as ADMX, CAST, and ALPS. Although star cooling rates constrain certain broad classes of ALPs \cite{Raf95,Raf99}, there are a number of loopholes in the astrophysical arguments (for example, they do not apply to spin-1 bosons) that permit, in principle, spin-spin interactions in the parameter space studied in the present work \cite{Raf99,Jai06}.

The most stringent constraints on exotic dipole-dipole interactions between electrons have been established by torsion-pendulum experiments \cite{Hec13,Ter15,Hun13} at the laboratory scale ($\lambda \gtrsim 1~{\rm cm}$) and by measurements on trapped ions \cite{Kot15} at the micron scale ($10~{\rm \mu m} \lesssim \lambda \lesssim 1~{\rm m}$). The only existing constraints on exotic dipole-dipole interactions between electrons at the atomic scale come from positronium spectroscopy \cite{Kar10,Les14,Kot15}, which carries a caveat that CPT invariance must be implicitly assumed in order to translate the constraint to electrons \cite{Kot15}. 

Spectroscopic measurements of helium have been a popular research topic for several decades \cite{Fen15, Pas04, Smi10, Zel05, Can99}. These investigations enable determination of energy-level structure of the element with a good precision. In particular, Feng \textit{et al.} recently \cite{Fen15} determined the frequency of the $2^3P_1 - 2^3P_2$ transition with an uncertainty of 0.36~kHz (1-$\sigma$ level) while measurements of the $2^3S_1 - 2^3P_{0,1,2}$ transitions performed by Pastor \textit{et al.}, measured the frequency with uncertainty of $\sim2$~kHz \cite{Pas04}.

To date, the most precise theoretical calculations of the helium energy structure have been performed by Pachucki and Yerokhin \cite{Pac10}, who used perturbation theory to calculate the helium fine-structure splittings up to the $m_e\alpha^7$ order (in relativistic units), where $m_e$ is the electron mass and $\alpha$ is the fine-structure constant. This enabled calculations of $2^3P_{0,1,2}$ level splittings with uncertainty of $\sim$2~kHz \cite{Pac10}. At the same time, the energy differences between the $2^3S_1$ and $2^3P_{0,1,2}$ levels were calculated to the $m_e\alpha^6$ order, enabling determination of the transition frequencies with uncertainties of $\sim$3.0~MHz \cite{Yer10}.

%In the context of comparison of experimental results and theoretical calculations, it is noteworthy that in quantum systems, e.g., atoms, a subtle systematic effect, arising from quantum interference of a given transition with off-resonant excitations, can affect positions of measured spectral lines \cite{Hor10, Hor11, Mar12, Mar14}. As reported in Ref. \cite{Mar14}, this effect causes the shift of the helium $2^3P_1 - 2^3P_2$ transition frequency by 10~kHz. Since this is larger than the uncertainty of experimental data and theoretical calculations, we use the value of the transition frequency reported in Ref. \cite{Fen15} where this effect was taken into account. On the other hand, the theoretical uncertainties of the $2^3S_1 - 2^3P_{0,1,2}$ transition frequencies are so large that the interference effect can be neglected in these cases.

In the context of comparison between experimental results and theoretical calculations of atomic energies, it is crucial to note a subtle systematic effect arising from quantum interference of a given atomic transition with off-resonant excitations can affect the resonant frequencies of measured spectral lines \cite{Hor10, Hor11, Mar12, Mar14}. For example, as discussed in detail in Ref. \cite{Mar14}, this effect can cause an apparent shift of the helium $2 ^3P_1 \rightarrow 2 ^3P_2$ transition frequency by $\approx 10~{\rm kHz}$.  These apparent shifts depend on the experimental technique used, since different techniques are sensitive to different quantum-mechanical interference paths \cite{Mar15}. Reference \cite{Mar15} summarizes the corrections and present status of measurements of the helium $2 ^3P_1 \rightarrow 2 ^3P_2$ interval, and we use the weighted average of the corrected results of Refs. \cite{Bor09, Fen15, Cas00, Zel05} to determine the experimental value for the $2 ^3P_1 \rightarrow 2 ^3P_2$ transition frequency. On the other hand, the theoretical uncertainties of the $2^3S_1 - 2^3P_{0,1,2}$ transition frequencies are so large that the interference effect can be neglected in these cases.

In this work, we determine limits on the coupling constants for various exotic interactions between electron spins from their possible effect on transition energies of helium. By comparing the experimental and theoretical results, we extract a maximal possible energy contribution $\Delta E$ that may come from exotic interactions at the $90\%$ confidence level (see Appendix A for details of how $\Delta E$ is determined). Table \ref{tab:1} presents the theoretical and experimental energy values for various $^4$He transitions used in our calculations of limits on exotic spin-dependent interactions. Note that the theoretical uncertainties are determined from estimates of the next-order contributions from quantum electrodynamics which are proportional to $m_e \alpha^8$, where $\alpha$ is the fine-structure constant.

\begin{table*}
\caption{\label{tab:1}Comparison of theoretical (QED-based) and experimental transition energies values between various helium states.}
\begin{ruledtabular}
\begin{tabular}{ccccccc}
% &\multicolumn{2}{c}{Theoretical} &\multicolumn{2}{c}{Experimental}& Difference & $\Delta E$ \\ \hline
 & Theoretical & & Experimental & & Difference & $\Delta E$ \\ \hline
$2^3 P_1 - 2^3 P_2$ & 2 291 178.9(1.7) kHz & \cite{Pac10} & 2 291 177.54(24) kHz & \cite{Mar15} & 1.4(1.7) kHz & 3.7 kHz\\
%$2^3 P_0 - 2^3 P_2$ & 31 908 131.2(1.8) kHz & \cite{Pac10} & 31 908 131.25(30) kHz & \cite{Smi10} & 0.1(1.8) kHz & 3.2 kHz \\
%$2^3 P_0 - 2^3 P_1$ & 29 616 952.3(1.7) kHz & \cite{Pac10} & 29 616 951.66(70) kHz & \cite{Zel05} & 0.6(1.8) kHz & 3.7 kHz\\ 
\hline
$2^3 P_0 - 2^3 S_1$ & 276 764 094.7(3.0) MHz & \cite{Yer10} & 276 764 094.7073(21) MHz & \cite{Pas04} & 0.0(3.0) MHz & 4.9MHz\\
$2^3 P_1 - 2^3 S_1$ & 276 734 477.7(3.0) MHz & \cite{Yer10} & 276 734 477.7525(20) MHz & \cite{Pas04} & 0.1(3.0) MHz & 5.0MHz\\
$2^3 P_2 - 2^3 S_1$ & 276 732 186.1(2.9) MHz & \cite{Yer10} & 276 732 186.621(15) MHz & \cite{Pas04} & 0.5(2.9) MHz & 5.3MHz\\
\end{tabular}
\end{ruledtabular}
\end{table*}

%\section{Spin-dependent potentials}
In Ref. \cite{Dob06}, Dobrescu and Mocioiu studied possible long-range potentials between fermions generated by exchange of spin-0 or spin-1 bosons. Given basic assumptions within the context of quantum field theory (e.g., rotational invariance, energy-momentum conservation, locality), interactions mediated by new bosons can generate sixteen independent, long-range potentials between fermions in the nonrelativistic limit (small fermion velocity and low momentum transfer). In the case of one-boson exchange under these assumptions, all the potentials acquire a dependence $\propto e^{-r_{12}/\lambda}$, where $r_{12}$ is the distance between the fermions, which largely determines the range of the exotic interactions. For example, the coupling of a pseudoscalar boson of mass $m_0$ to an electron $\psi$ can arise as either a Yukawa-like coupling described by the Lagrangian \cite{Moo84}
\begin{align}
\mathcal{L}_{Yuk} = -i g_p \bar{\psi} \gamma^5 \psi \varphi~,
\label{Eq:V3-Yukawa-Lagrangian}
\end{align}
or through a derivative coupling described by the Lagrangian
\begin{align}
\mathcal{L}_{Der} = \frac{g_p}{2m_e} \bar{\psi} \gamma_\mu\gamma^5 \psi \partial^\mu \varphi~,
\label{Eq:V3-derivative-Lagrangian}
\end{align}
where in Eqs.~\eqref{Eq:V3-Yukawa-Lagrangian} and \eqref{Eq:V3-derivative-Lagrangian} we have used the Dirac $\gamma$ matrices. In either case, it turns out that the resultant long-range potential is given by:
\begin{widetext}
\begin{align}
V_3=\frac{g_3^e g_3^e}{4\pi\hbar c} \frac{\hbar^3}{4 m_e^2 c} \left[\textbf{s}_1\cdot\textbf{s}_2\left(\frac{1}{\lambda r_{12}^2}+\frac{1}{r_{12}^3}\right)-\left(\textbf{s}_1\cdot\textbf{e}_{12}\right)\left(\textbf{s}_2\cdot\textbf{e}_{12}\right)\left(\frac{1}{\lambda^2 r_{12}}+\frac{3}{\lambda r_{12}^2}+\frac{3}{r_{12}^3}\right)\right]e^{-r_{12}/\lambda},\label{eq:v3}
\end{align}
\end{widetext}
where $g^e_i g^e_i/(4\pi \hbar c)$ is the dimensionless coupling constant of the $i$-th interaction between the electrons (this is the notation of Refs. \cite{Dob06, Kot15, Jac10, Moo84}, where $g^e$ refers to the coupling of an electron to the exotic boson), $m_e$ is the electron mass, $\textbf{e}_{12}=\textbf{r}_{12}/r_{12}$ is the unit vector in the direction from the first electron to the second electron, $\nabla_1$ and $\nabla_2$ are vector differential operators in position space for the first and second particle respectively, and  $\textbf{s}_1$, $\textbf{s}_2$ are spins of the interacting electrons. Further details of the derivation of the long-range spin-dependent potentials are given in Refs. \cite{Moo84, Dob06} and Appendix B. For studies of exotic spin couplings using $^4$He, only those potentials invariant under permutation of identical fermions, spatial inversion, and time reversal are relevant. These three conditions allow a non-zero result of calculations of exotic-field-induced shifts of energy levels in first-order perturbation theory. There are four potentials that satisfy these requirements. One of them was introduced in Eq. (\ref{eq:v3}), and the other three have in the position representation the form
\begin{widetext}
\begin{eqnarray}
		V_2&=&\frac{g_2^e g_2^e}{4\pi\hbar c} \hbar c \left(\textbf{s}_1\cdot\textbf{s}_2\right) \frac{e^{-r_{12}/\lambda}}{r_{12}},\label{eq:v2}\\
		%V_3&=&\frac{g_3^e g_3^e}{4\pi\hbar c} \frac{\hbar^3}{4 m_e^2 c} \left[\textbf{s}_1\cdot\textbf{s}_2\left(\frac{1}{\lambda r_{12}^2}+\frac{1}{r_{12}^3}\right)-\left(\textbf{s}_1\cdot\textbf{e}_{12}\right)\left(\textbf{s}_2\cdot\textbf{e}_{12}\right)\left(\frac{1}{\lambda^2 r_{12}}+\frac{3}{\lambda r_{12}^2}+\frac{3}{r_{12}^3}\right)\right]e^{-r_{12}/\lambda},\label{eq:v3}\\
		%V_{4}&=& \frac{g_4^e g_4^e}{4\pi\hbar c} \frac{i \hbar^3}{4 m_e^2 c} (\textbf{s}_1 + \textbf{s}_2) \cdot \left[(\nabla_1-\nabla_2)\times\textbf{e}_{12}\left(\frac{1}{r_{12}^2}+\frac{1}{\lambda r_{12}}\right)e^{-r_{12}/\lambda}-e^{-r_{12}/\lambda}\left(\frac{1}{r_{12}^2}+\frac{1}{\lambda r_{12}}\right)\textbf{e}_{12}\times(\nabla_1-\nabla_2)\right],\label{eq:v4}\\
		V_{4}&=& \frac{g_4^e g_4^e}{4\pi\hbar c} \frac{i \hbar^3}{4 m_e^2 c} (\textbf{s}_1 + \textbf{s}_2) \cdot \left[(\nabla_1-\nabla_2)\times\textbf{r}_{12},\left(\frac{1}{r_{12}^3}+\frac{1}{\lambda r_{12}^2}\right)e^{-r_{12}/\lambda}\right]_{+},\label{eq:v4}\\
		V_8&=&\frac{g_8^e g_8^e}{4\pi\hbar c} \frac{\hbar^3}{4 m_e^2 c} \left[\textbf{s}_1\cdot(\nabla_1-\nabla_2),\left[\textbf{s}_2\cdot(\nabla_1-\nabla_2), \frac{e^{-r_{12}/\lambda}}{r_{12}}\right]_{+}\right]_{+},\label{eq:v8}
\end{eqnarray}
\end{widetext}
where by $[\cdot,\cdot]_{+}$ we denote an anticommutator. These potentials are results of the exchange of exotic bosons \cite{Dob06, Kar11, Les14}: scalar ($V_4$), pseudoscalar ($V_3$), vector ($V_3$), and axial-vector ($V_2, V_3, V_8$).

Note that the velocity-dependent potentials [Eqs. (\ref{eq:v4}) and (\ref{eq:v8})] presented here have different forms than in Ref. \cite{Dob06} and other papers considering non-static exotic interactions \cite{Hun14}. This difference comes from the fact that the velocity-dependent potentials in Refs. \cite{Dob06, Hun14} are in fact presented in a ``mixed'' representation (not a position representation, as stated). We discuss this further in the Appendix B.

%\section{Helium wave functions}
The strength of any hypothetical exotic spin-dependent interactions between two electrons is orders of magnitude smaller than their electromagnetic interaction. Based on this fact, high precision is not required in calculation of the perturbation due to the exotic effects and it is enough to calculate the exotic contributions to first order in perturbation theory. For these calculations, approximate wave functions of electrons in helium may be assumed. Here, we use the electron wave functions of the $n=2$ state of orthohelium ($S=1$), obtained with the variational method (see, for example, Ref. \cite{Bethe}). In Table \ref{tab:2}, one can find the ionization energies calculated with these wave functions $\varepsilon_{th}$ compared with the experimental values $\varepsilon_{exp}$. The difference between them is just several percent, which suggests that these functions can be safely used in our calculations. These approximate wave functions have reasonable accuracy only for distances on the order of the Bohr radius $a_0$. Larger distances do not contribute in our estimates because all potentials decrease faster than $1/r$. Shorter distances $r \ll a_0$ can be important for singular potentials, but the potentials we consider are not singular for the boson masses $m_0 \lesssim 1~{\rm keV}$ studied here and consequently their matrix elements mainly depend on the distances $r \sim a_0$. For $m_0 \gg 1~{\rm keV}$ the potentials become singular and the accuracy of our estimates decrease accordingly, but this is exactly the regime in which the strength of our constraints decrease. Thus, these approximate wave functions are adequate for the part of the parameter space where our constraints are significant.

The spatial electron wave function for the helium $2^3S_0$  state is given by \cite{Bethe}
\begin{eqnarray}
		\psi^S=C^S\left[e^{-Z_i^S r_1/a_0 - Z_a^S r_2/2a_0}\left (\frac{Z_a^S r_2}{2a_0} -1\right)\right.-\nonumber\\
				\left. - e^{-Z_i^S r_2/a_0 - Z_a^S r_1/2a_0}\left (\frac{Z_a^S r_1}{2a_0} -1\right)\right],\label{eq:23S}
\end{eqnarray}
where the $Z_a^S, Z_i^S, C^S$ values are given in Table \ref{tab:2} and $a_0$ is the Bohr radius. The spatial electron wave function is antisymmetric with respect to the $1 \leftrightarrow 2$ electron exchange, so the spin wave function must be symmetric (as we may expect for orthohelium) and the total spin is $S=1$. Since the $2^3S_0$ state is only used to constrain the $V_2$ potential, where the electron spins appear in the formula via the $\textbf{s}_1\cdot \textbf{s}_2$ term, we do not have to consider explicitly the spin part of the wave function as for orthohelium
\begin{equation}
\textbf{s}_1\cdot\textbf{s}_2 |\psi^S\rangle=\frac{1}{2} (\textbf{S}^2-\textbf{s}_1^2-\textbf{s}_2^2)|\psi^S\rangle=\frac{1}{4}|\psi^S\rangle.
\end{equation}

The spatial components of the $2^3P$-state wave functions are approximated by \cite{Bethe}
\begin{eqnarray}
		\Psi^P_1&=&-C^P[F(r_1,r_2)\sin\theta_1 e^{i\phi_1}-F(r_2,r_1)\sin\theta_2 e^{i\phi_2}],	\nonumber \\
		\Psi^P_0&=&\sqrt{2}C^P[F(r_1,r_2)\cos\theta_1-F(r_2,r_1)\cos\theta_2], \nonumber\\
		\Psi^P_{-1}&=&C^P[F(r_1,r_2)\sin\theta_1 e^{-i\phi_1}-F(r_2,r_1)\sin\theta_2 e^{-i\phi_2}],\nonumber\\
		\label{eq:P1}
\end{eqnarray}
where
\begin{equation}
F(r_1,r_2)=\frac{r_1}{a_0} e^{-Z_a^P r_1/2a_0-Z_i^P r_2/a_0},
\end{equation}
where the $Z_a^P, Z_i^P, C^P$ values are given in Table \ref{tab:2}. We associate these antisymmetric wave functions with symmetric spin functions using the Clebsch-Gordan coefficients coming from addition of angular momenta $L = 1$ and $S = 1$. In the following sections, we will be performing calculations using wave functions $|\psi^P_{J,m_J}\rangle$
\begin{eqnarray}
		|\Psi^P_{2,2}\rangle&=&\Psi^P_1|\uparrow\uparrow\rangle,	\label{eq:23P22}\\
		|\Psi^P_{2,1}\rangle&=&\sqrt{\frac{1}{2}}\Psi^P_0|\uparrow\uparrow\rangle+\frac{1}{2}\Psi^P_1\left(|\uparrow\downarrow\rangle+|\downarrow\uparrow\rangle\right),	\label{eq:23P21}
\end{eqnarray}
where $|\uparrow\downarrow\rangle=|m_{s_1}=1/2; m_{s_2}=-1/2\rangle$ and $m_{s_{1,2}}$ are the magnetic quantum numbers of the 1$^{st}$ and 2$^{nd}$ electron, respectively.

\begin{table}
\caption{\label{tab:2}Values of constants in the wave functions and ionization energies.}
\begin{ruledtabular}
\begin{tabular}{cccccc}
 & $Z_i$ & $Z_a$ & $C$ & $\varepsilon_{th}$ & $\varepsilon_{exp}$ \\ \hline
 $2^3 S$ & 2.01 & 1.53 & 0.43247 $a_0^{-3/2}$ & 0.334 Ry & 0.350 Ry \\
 $2^3 P$ & 1.99 & 1.09 & 0.097969 $a_0^{-3/2}$ & 0.262 Ry & 0.266 Ry \\
\end{tabular}
\end{ruledtabular}
\end{table}

%\section{Results}\label{sec:results}
For every considered potential $V_i$ we can  estimate an associated energy shift between states $|\psi_a\rangle$ and $|\psi_b\rangle$ using first-order perturbation theory and the approximate wave functions listed above
\begin{equation}
\Delta U_{ab,i}(m_0)=\langle \psi_a| \mathcal{V}_i(m_0) |\psi_a\rangle - \langle \psi_b| \mathcal{V}_i(m_0) |\psi_b\rangle,
\label{eq:Uabi}
\end{equation}
where $\mathcal{V}_i(m_0)$ is the potential $V_i$ divided by the dimensionless constant $g^e_i g^e_i/(4\pi \hbar c)$. Values for $\Delta U_{ab,i}$ were calculated by numerical integration for several $m_0$ values and then an interpolation was performed in order to obtain a continuous function $\Delta U_{ab,i}(m_0)$. For potentials $V_3, V_4$, and $V_8$ curves describing the constraints on $g^e_i g^e_i/(4\pi \hbar c)$ were obtained for different values of $m_0$ by substituting the appropriate $\Delta E$ (the one connected with the $2^3 P_1 - 2^3 P_2 $ transition) from Table \ref{tab:1} into the relation:
\begin{equation}
\frac{g^e_i g^e_i}{4\pi \hbar c}(m_0)\leq\frac{\Delta E}{\Delta U_{ab,i}(m_0)}.
\label{eq:gege}
\end{equation}

For $m_0\gtrsim3000$ eV, the Compton wavelength of the mediating boson is shorter than the average interparticle separation between electrons in the helium atom. Because of that, the transition frequency becomes less sensitive to the considered potentials for $m_0 \gtrsim 3000$ eV as seen in the parameter exclusion plots.

\begin{figure}
\includegraphics[width=0.45\textwidth]{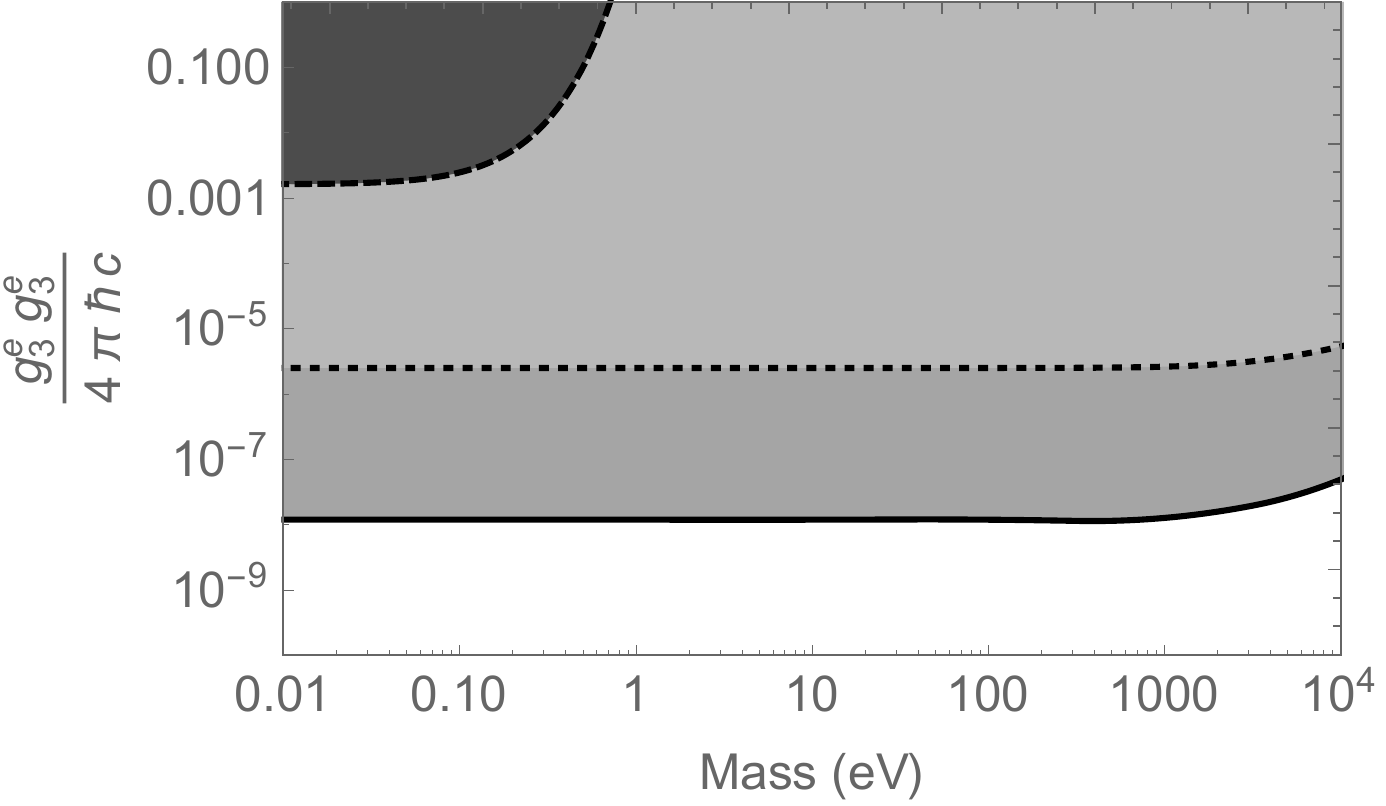}\\
\caption{\label{fig:v3}Constraints (at the 90$\%$ confidence level) on the dimensionless coupling constant $g^e_3 g^e_3/(4\pi \hbar c)$ as a function of the boson mass. The dashed line and dark gray fill shows the constraint for electrons from Ref. \cite{Kot15}. The dotted line and light gray fill show the constraint derived from analysis of positronium, also discussed in \cite{Kot15}. The solid line and medium gray fill shows the constraint from a comparison between theory and experiment for the $2 ^3P_2 - 2 ^3P_1$ transition frequency in He.}
\end{figure}

%\subsection{Potential $V_{3}$}
%We use the first-order perturbation theory to calculate constraints for the $V_3$ potential from Eqs. (\ref{eq:Uabi}), (\ref{eq:gege}) using the $2^3 P_2$ and $2^3 P_1$ states' wave functions given by Eqs. (\ref{eq:23P22}), (\ref{eq:23P21}). We test a range of $\lambda$ corresponding to boson masses of $10^{-2}$--$10^4$ eV.
The results for the $V_3$ potential are presented in Fig.~\ref{fig:v3}. The other results in this figure come from Ref. \cite{Kot15}. It can be seen that comparison between theory and experiment for helium fine structure yields the best constraints in the considered mass range (two orders of magnitude more stringent than the previous limits).

\begin{figure}
\includegraphics[width=0.45\textwidth]{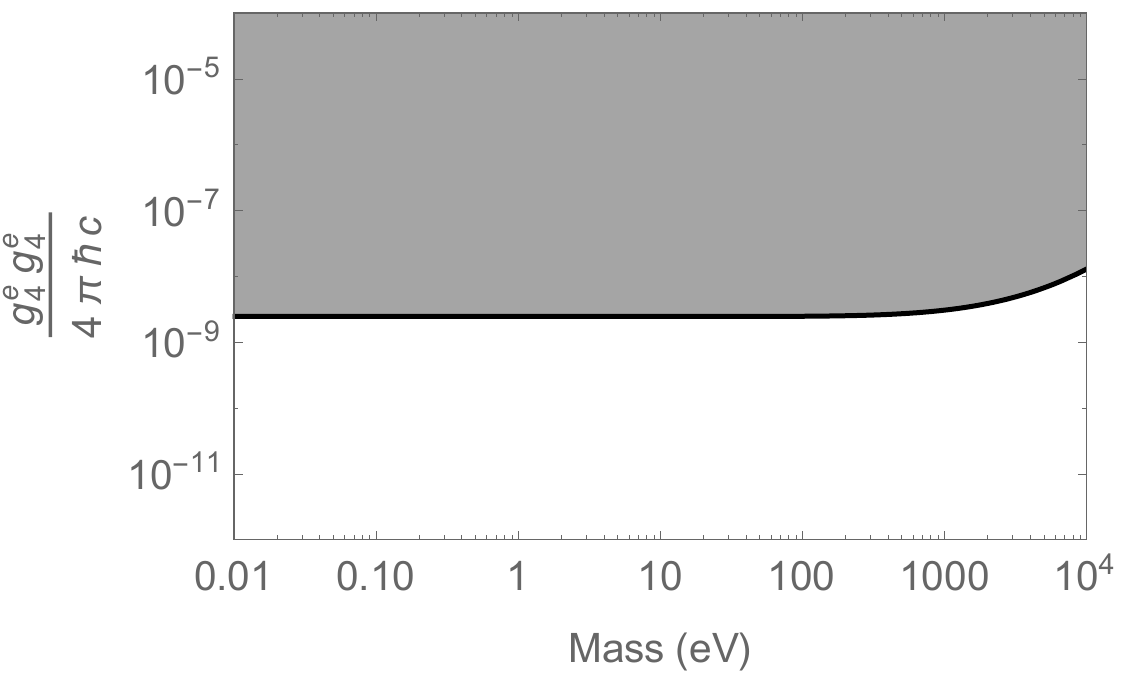}
\caption{\label{fig:v4}Constraints (at the 90$\%$ confidence level) on the dimensionless coupling constant $g^e_4 g^e_4/(4\pi \hbar c)$ as a function of the boson mass coming from a comparison between theory and experiment for the $2 ^3P_2 - 2 ^3P_1$ transition frequency in He.}
\end{figure}

%\subsection{Potential $V_{4}$}\label{sec:v4}
In order to calculate constraints for the $V_4$ potential we use its reduced form, which is derived in the Appenidx C. It allows us to numerically obtain $\Delta U_{ab,i}(m_0)$ function plotted in Fig. \ref{fig:v4}.

\begin{figure}
\includegraphics[width=0.45\textwidth]{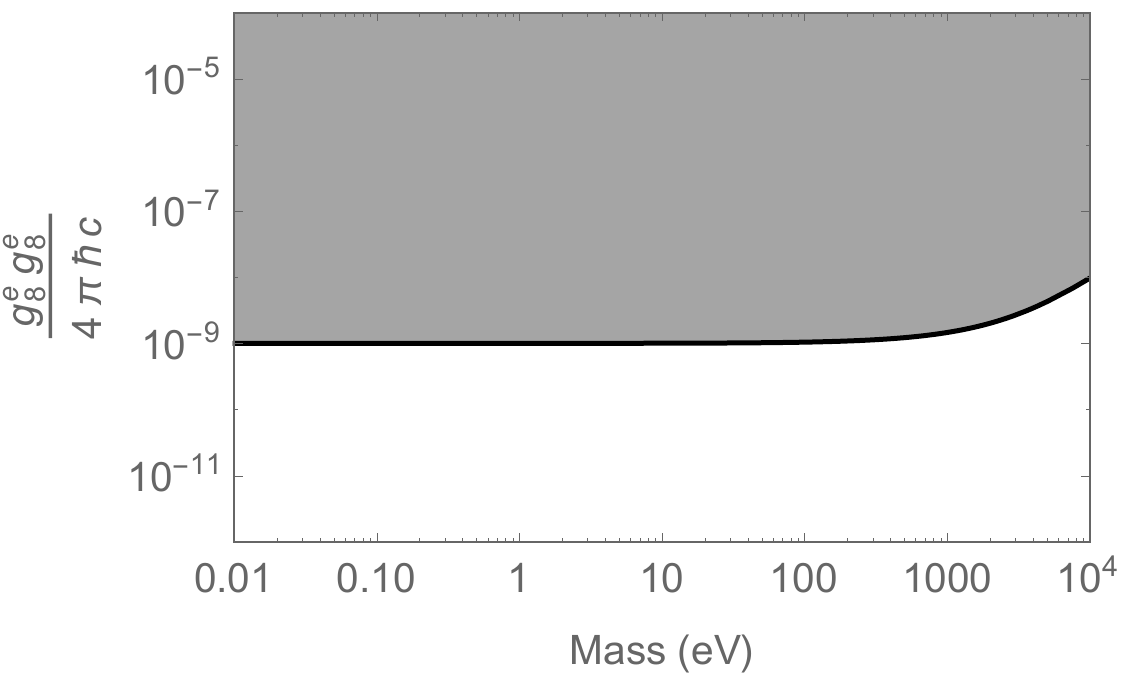}
\caption{\label{fig:v8}Constraints (at the 90$\%$ confidence level) on the dimensionless coupling constant $g^e_8 g^e_8/(4\pi \hbar c)$ as a function of the boson mass coming from a comparison between theory and experiment for the $2 ^3P_2 - 2 ^3P_1$ transition frequency in He.}
\end{figure}

%\subsection{Potential $V_{8}$}
The results for the $V_8$ potential are presented in Fig.~\ref{fig:v8}. Constraints for $V_8$ electron coupling constant were obtained earlier using geoelectron experiments \cite{Hun14}, which considered boson masses less than $10^{-10}$ eV, yielding constraints $g^e_8 g^e_8/(4\pi \hbar c)\lesssim 10^{-36}$ in the massless limit.

%\subsection{Potential $V_{2}$}\label{sec:v2}
The analysis for the $V_2$ potential differs somewhat from that carried out for the other potentials. Spin operators in the $V_2$ potential are of the form $\textbf{s}_1\cdot\textbf{s}_2$ so  for orthohelium wave functions $|\psi\rangle$ we have $\textbf{s}_1\cdot\textbf{s}_2 |\psi\rangle=\tfrac{1}{4}|\psi\rangle$. This means that the analysis for this case is based on evaluation of the $\langle \psi | \exp(-r_{12}/\lambda)/r_{12} | \psi \rangle$ matrix elements.

\begin{figure}
\includegraphics[width=0.45\textwidth]{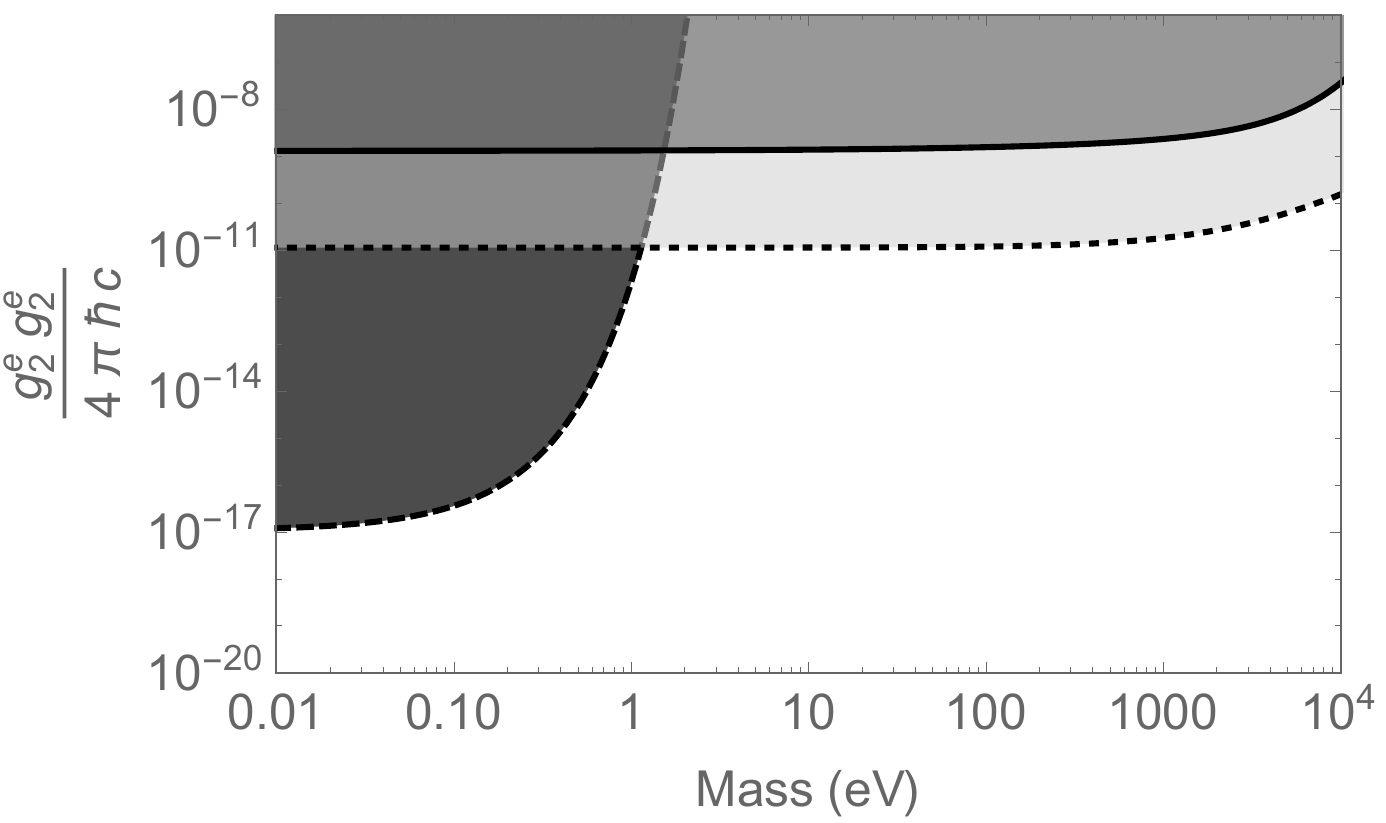}\\
\caption{\label{fig:v2}Constraints (at the 90$\%$ confidence level) on the dimensionless coupling constant $g^e_2 g^e_2/(4\pi \hbar c)$ as a function of the boson mass. The dashed line and dark gray fill shows the constraint for electrons from Ref. \cite{Kot15}. The dotted line and light gray fill show the constraint derived from analysis of positronium, also discussed in \cite{Kot15}. The solid line and medium gray fill shows the constraint from a comparison between theory and experiment for the $2 ^3S_1 - 2 ^3P$ transition frequency in He.}
\end{figure}

The $V_2$ potential does not split energy levels of different $J$ and the same $L$ and $S$, but only shifts such levels by the same amount. This means that in order to experimentally observe the shifts, we need another reference state outside the fine-structure manifold. For this purpose, based on the available experimental data and theoretical calculations, a natural choice is a comparison between the $2^3 S_1$ and $2^3 P$ states. The fact that the $V_2$ potential does not remove $J$ degeneracy implies that the $2^3 S_1 - 2^3 P_J$ comparison does not depend on the particular choice of $|J m_J\rangle$. Therefore, we use all the values of differences between experimental and theoretical transition energies between states $2^3 S$ and $2^3 P$ from Table \ref{tab:1}. Treating these differences as $\Delta E$ from formula (\ref{eq:gege}), we get a function $g^e_2 g^e_2/(4\pi \hbar c)(m_0)$ for every transition, along with the uncertainty. We calculate weighted mean of these with its uncertainty, and take a sum of this mean and a doubled uncertainty as the limit. The results are presented in Fig. \ref{fig:v2}. The obtained constraints are worse than the ones obtained using positronium \cite{Kot15}, but we note that positronium constrains the interaction between positrons and electrons which can only be directly compared with the electron-electron interaction under the assumption of CPT invariance.

%\section{Summary and Outlook}
In conclusion, by comparing the results of precision spectroscopic measurements in $^4$He with theoretical calculations of the corresponding energy intervals, we establish constraints on possible exotic interactions that could arise due to the exchange of bosonic fields, as introduced in the theoretical framework of Refs. \cite{Moo84, Dob06}. We point out an inconsistency of the operator definitions in Ref.~\cite{Dob06} and perform the analysis with the corrected operators. We improve constraints on the strength of some of the exotic interactions by two orders of magnitude and constrain others for the first time.

We expect He spectroscopy to become an even more sensitive probe of exotic electron-electron interactions as atomic theory and experiment become more precise.

\begin{acknowledgments}
The authors acknowledge Victor Flambaum, Krzysztof Pachucki and Savely G. Karshenboim for fruitful discussions and useful remarks. This project was partially supported by the European Research Council (ERC) under the European Union’s Horizon 2020 research and innovation programme (grant agreement No 695405), Russian Foundation for Basic Research Grant No.\ 14-02-00241, the Polish Ministry of Science and Higher Education within the Iuventus Plus program (grant 0390/IP3/2015/73) and the Diamond Grant (grant 0143/DIA/2016/45), National Science Foundation under grant PHY-1307507, the Heising-Simons and Simons Foundations, and a Marie Curie International Incoming Fellowship within the 7th European Community Framework Programme. MK is grateful to Mainz Institute for Theoretical Physics (MITP) for its hospitality and support.
%This project has received partial funding from the European Research Council (ERC) under the European Union’s Horizon 2020 research and innovation programme (grant agreement No 695405).  This work is partly supported by Russian Foundation for Basic Research Grant No.\ 14-02-00241. MK is grateful to Mainz Institute for Theoretical Physics (MITP) for its hospitality and support.

\end{acknowledgments}

\appendix

\section{Analysis of the experimental and theoretical data}
Determination of the constraints on the exotic spin-dependent interactions between two electrons in \mbox{helium-4} requires comparison of experimental and theoretical data. Particularly, uncertainty of the data needs to be considered to constraint such interactions at a given acceptance level (here, 90\%). In our approach, for fine-structure considerations, where we use only one transition, the quantity $\Delta E$ is given by 
\begin{equation}
\Delta E = \textrm{max}\{|\mu + L|, |\mu-L|\},
\end{equation}
where $\mu$ is the mean difference between theoretical and experimental transition energies and $L$ is determined in such a way that
\begin{equation}
0.9=\int_{-L}^{+L}\frac{1}{\sqrt{2\pi}\sigma}e^{-(x-\mu)^2/(2\sigma^2)}dx,
\end{equation}
where $\sigma$ is the resultant uncertainty, originating from theoretical ($\sigma_{th}$) and experimental ($\sigma_{exp}$) uncertainties combined in quadrature, $\sigma^2=\sigma^2_{th}+\sigma^2_{exp}$. This method was used for the potentials $V_3, V_4$, and $V_8$.

In the case of $V_2$ potential, we use all the values of differences between experimental and theoretical transition energies between states $2^3 S$ and $2^3 P$ as $\Delta E$, as described in the paper. We obtain a function $g^e_2 g^e_2/(4\pi \hbar c)(m_0)$, along with the uncertainty, separately for every transition. We calculate weighted mean of these with its uncertainty, and take a sum of this mean and a doubled uncertainty as the limit. This sum is our final constraint $g^e_2 g^e_2/(4\pi \hbar c)(m_0)$.

It should be noted that the method used to determined constraints for the $g_2^e g_2^e/(4\pi\hbar c)$ could be also used to determine constraints for the potentials $V_3$, $V_4$, and $V_8$. In fact constraints obtained this way are twice more stringent than the ones plotted in Figs. 1-3 of the paper, however, we do not use them, as they include a systematic error due to the shifts from a distant neighboring resonance (see discussion in the article).

\section{Potentials in position representation}
The interaction potentials presented in the paper [Eqs. (\ref{eq:v3})-(\ref{eq:v8})] differ from their counterparts presented in Ref. \cite{Dob06}. Here, we show derivation of the potentials used for our calculations and explain the source of the difference between the ones presented in Ref. \cite{Dob06}.

\begin{figure}
\includegraphics[width=0.45\textwidth]{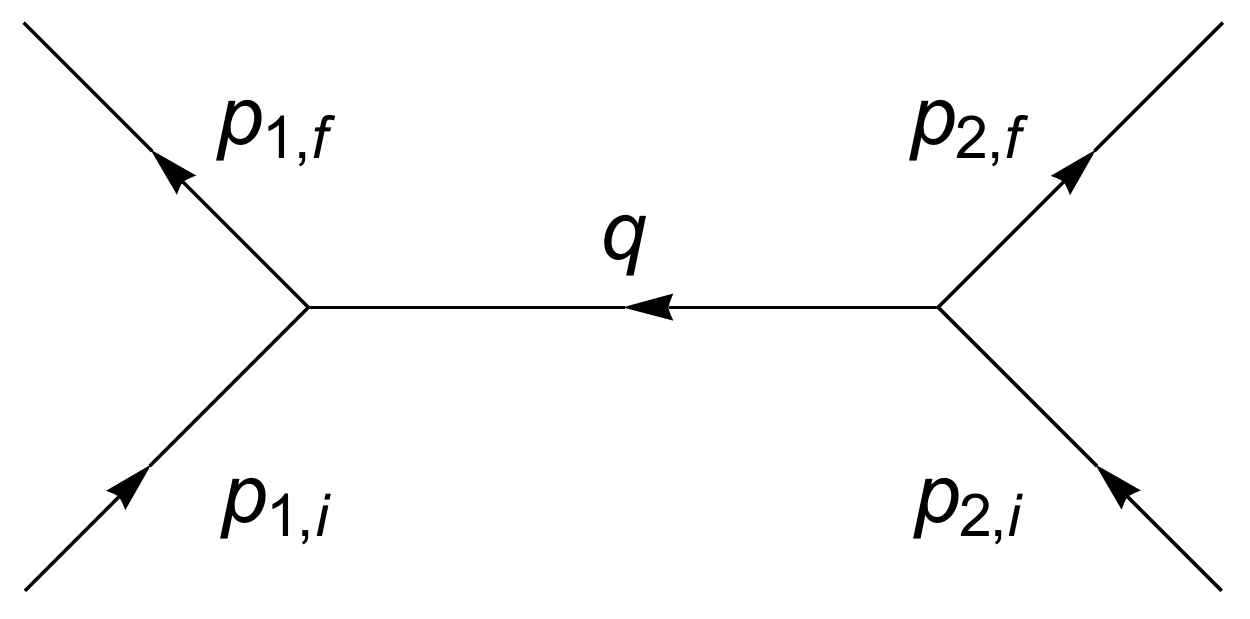}\\
\caption{\label{fig:graph} Feynman diagram of an interaction between two electrons mediated by a light boson.}
\end{figure}

Let us consider an interaction between two electrons mediated by a light boson. A corresponding Feynman diagram is shown in Fig. \ref{fig:graph}, where $\textbf{p}_{1,i}$ and $\textbf{p}_{1,f}$ are initial and final momenta of the first electron ($\textbf{p}_{2,i}$ and $\textbf{p}_{2,f}$ are analogously initial and final momenta of the second electron) and $\textbf{q}$ is the momentum of the light interacting boson. We may describe this interaction in the center of mass frame using just two vectors
\begin{eqnarray}
\textbf{P}&=&\tfrac{1}{2}\left(\textbf{p}_{1,f}+\textbf{p}_{1,i}\right),\\
\textbf{q}&=&\textbf{p}_{1,f}-\textbf{p}_{1,i}.
\end{eqnarray}
In their paper \cite{Dob06}, Dobrescu and Mocioiu construct 16 independent, rotationally invariant scalars consisting of the vectors $\textbf{P}, \textbf{q}, \textbf{s}_1, \textbf{s}_2$, where $\textbf{s}_1, \textbf{s}_2$ are the spin of the first and second electron, respectively. These scalars are operators in momentum representation (momentum operators are multiplication operators). Due to the focus of this paper, we consider only four of them that are spin-dependent, symmetric with respect to a permutation of identical fermions, and invariant under spatial inversion and time reversal. In natural units ($c=\hbar=1$), they take forms
\begin{eqnarray}
		\mathcal{O}_2&=&\textbf{s}_1 \cdot \textbf{s}_2,\label{eq:o2}\\
		\mathcal{O}_3&=&\frac{1}{m_e^2} \left(\textbf{s}_1\cdot\textbf{q}\right)  \left(\textbf{s}_2\cdot\textbf{q}\right),\label{eq:o3}\\
		\mathcal{O}_4&=&\frac{i}{2m_e^2} \left(\textbf{s}_1 + \textbf{s}_2\right) \cdot \left(\textbf{P}\times\textbf{q}\right),\label{eq:o4}\\
		\mathcal{O}_8&=&\frac{1}{m_e^2} \left(\textbf{s}_1\cdot\textbf{P}\right)  \left(\textbf{s}_2\cdot\textbf{P}\right)\label{eq:o8},
\end{eqnarray}
where $m_e$ is an electron mass.  Note that $i\textbf{q}$, rather than $\textbf{q}$, is a Hermitian operator, which is why the ${\cal O}_4$ operator [Eq.(\ref{eq:o4})], linear in $\textbf{q}$, is imaginary.

In Sec.\ 3 of Ref. \cite{Dob06}, the operators $\mathcal{O}_2$, $\mathcal{O}_3$, $\mathcal{O}_4$, and $\mathcal{O}_8$ are converted into potentials by making a Fourier transform from $\textbf{q}$ to $\textbf{r}_{12}= \textbf{r}_1 -\textbf{r}_2$ (we introduce here a slightly different notation than the original one). Note that this is a mixed representation as the authors still keep $\textbf{v} = \textbf{P}/m_e$ as a variable, rather than an operator (see Eq.\ (3.2) in Ref. \cite{Dob06}). In the position representation, all expressions which include $\textbf{v}$ should be written in terms of an operator $\hat{\textbf{v}}$, which is related to a gradient.

Let us consider a potential of the form $ \textbf{P} V(r_{12})$
\begin{eqnarray}\label{eq:PV_in_r_rep}
\quad&\quad&\langle \psi_f(r_1,r_2)| \textbf{P} V(r_{12})|\psi_i(r_1,r_2)\rangle\nonumber \\
\quad&=& \tfrac12 \langle \psi_f(r_1,r_2)|\textbf{p}_{1,f} V(r_{12})+ V(r_{12})\textbf{p}_{1,i}|\psi_i(r_1,r_2)\rangle\nonumber \\
\quad&=& \tfrac12 \langle \psi_f(r_1,r_2)|\hat{\textbf{p}}_{1} V(r_{12})+ V(r_{12})\hat{\textbf{p}}_{1}|\psi_i(r_1,r_2)\rangle\nonumber\\
\quad&=& \tfrac12 \left\langle \psi_f(r_1,r_2)\left|\left[\hat{\textbf{p}}_1, V(r_{12})\right]_+\right|\psi_i(r_1,r_2)\right\rangle,
\end{eqnarray}
where $|\psi_i(r_1,r_2)\rangle$ and $|\psi_f(r_1,r_2)\rangle$ are the initial and final states of the considered system, respectively, and $\hat{\textbf{p}}_1$ is the momentum operator of the first electron. Omitting this step, as in Ref. \cite{Dob06}, results in mixed representation of non-static potentials, where $\textbf{v}$ is a variable rather than an operator.

Having this in mind, we perform the Fourier transform in order to go from the momentum representation of the potentials to their position representation:
\begin{equation}
\widetilde{V}_i(\textbf{r}_{12},\textbf{p}_{12})=-\int \frac{d^3 q}{(2\pi)^3} e^{i\textbf{q}\textbf{r}_{12}}\mathcal{P}(\textbf{q}^2)\mathcal{O}_i(\textbf{q},\textbf{P}),
\end{equation}
\\
where $\mathcal{P}(\textbf{q}^2)$ is a propagator. We are interested in Lorentz invariant exotic interactions communicated by a single boson with mass $m_0$, which implies a propagator of the form \cite{Dob06, Berestetskii}
\begin{equation}
	\mathcal{P}(\textbf{q}^2)=-\frac{1}{\textbf{q}^2+m_0^2}.
\end{equation}
Useful formulae for these Fourier transforms may be found in Appendix B of Ref. \cite{Dob06}. 

As an example we will derive the position--representation form of the $V_4$ potential. We begin with the momentum--representation form in the natural units [Eq. (\ref{eq:o4})]. By performing Fourier transform we obtain
\begin{eqnarray}
\label{eq:fourier}
\widetilde{V}_4&=&\int \frac{d^3 q}{(2\pi)^3} e^{i\textbf{q}\textbf{r}_{12}} \frac{ \mathcal{O}_4}{\textbf{q}^2+m_0^2}\\
&=&\frac{i}{2m_e^2} \left(\textbf{s}_1 + \textbf{s}_2\right)\cdot\left( \textbf{P}\times \int \frac{d^3 q}{(2\pi)^3} e^{i\textbf{q}\textbf{r}_{12}} \frac{\textbf{q}}{\textbf{q}^2+m_0^2}\right)\nonumber\\
&=&-\frac{1}{8\pi m^2_e} \left(\textbf{s}_1 + \textbf{s}_2\right)\cdot\left( \textbf{P}\times\frac{\textbf{r}_{12}}{r^3_{12}}\right) \left(1+m_0 r_{12}\right) e^{-m_0 r_{12}}.\nonumber
\end{eqnarray}

Now let us apply similar reasoning as in case of Eq.~(\ref{eq:PV_in_r_rep}), but for the operator $ \textbf{P}\times\textbf{r}_{12} V(r_{12})$. The $j$-th component of this operator matrix element will be (using the Einstein summation convention):
\begin{widetext}
\begin{eqnarray*}
 \left( \langle \psi_f(r_1,r_2)|\textbf{P}\times\textbf{r}_{12} V(r_{12})|\psi_i(r_1,r_2)\rangle\right)_j 
 = \left(\tfrac12 \langle \psi_f(r_1,r_2)|(\textbf{p}_{1,f}+\textbf{p}_{1,i}) \times (\textbf{r}_{12} V(r_{12}))|\psi_i(r_1,r_2)\rangle \right)_j\\
 = \tfrac12 \varepsilon_{jkl}\langle \psi_f(r_1,r_2)|(p_{1,f}^k+p_{1,i}^k) r_{12}^l V(r_{12})|\psi_i(r_1,r_2)\rangle 
 = \tfrac12 \varepsilon_{jkl} \left\langle \psi_f(r_1,r_2)\left| \left(p_{1}^k r_{12}^l V(r_{12})+ V(r_{12}) r_{12}^l p_{1}^k\right) \right|\psi_i(r_1,r_2)\right\rangle \\
 = \tfrac12 \varepsilon_{jkl} \left\langle \psi_f(r_1,r_2)\left| \left(p_{1}^k r_{12}^l V(r_{12})+ V(r_{12}) p_{1}^k r_{12}^l \right) \right|\psi_i(r_1,r_2)\right\rangle 
= \tfrac12  \varepsilon_{jkl}\langle \psi_f(r_1,r_2)|\left[ p_{1}^k  r_{12}^l , V(r_{12})\right]_{+} |\psi_i(r_1,r_2)\rangle \\
= \tfrac12  \langle \psi_f(r_1,r_2)|\left[ (\textbf{p}_{1} \times \textbf{r}_{12})_j , V(r_{12})\right]_{+} |\psi_i(r_1,r_2)\rangle,
\end{eqnarray*}
\\
\end{widetext}
where we have used the fact that $\varepsilon_{jkl}r^k p^l=\varepsilon_{jkl} p^l r^k+i \varepsilon_{jkl} \delta^{kl}=\varepsilon_{jkl} p^l r^k$. These calculations were performed in the center of mass frame of the two particles. We can convert obtained equations to the atom center of mass frame by substituting $\textbf{p}_1 \to \textbf{p}_1 - \tfrac{1}{2} (\textbf{p}_1+\textbf{p}_2)=\tfrac{1}{2} (\textbf{p}_1-\textbf{p}_2)=\tfrac{1}{2}\textbf{p}_{12}$. When we insert results of these calculations into Eq. (\ref{eq:fourier}), we get
\begin{equation*}
\widetilde{V}_4=-\frac{1}{16\pi m^2_e} \left(\textbf{s}_1 + \textbf{s}_2\right)\cdot\left[ \textbf{p}_{12}\times\textbf{r}_{12}, \frac{1+m_0 r_{12}}{r^3_{12}} e^{-m_0 r_{12}}\right]_{+}.
\end{equation*}
This form of the potential is used to calculate the contribution of the $V_4$ interaction to the helium energy levels. The last remaining steps are introducing the coupling constant, writing momenta as a differential operators, and inserting physical constants: $c$, $\hbar$, $m_e$, and the reduced Compton wavelength of the interaction boson $\lambda=\hbar/m_0 c$. These steps result in Eq. (\ref{eq:v4}) of the paper.

As a final remark. The framework introduced to deal with exotic potentials by Dobrescu and Mocioiu in Ref. \cite{Dob06} works only in the low-mass limit of the interacting boson. However, as we are interested in bosons with atomic-scale Compton wavelength, we can safely treat this framework as accurate.

\section{The reduced form of $V_4$ potential}
Performing numerical calculation of $V_4$ matrix elements is tedious due to the potential's complexity. However, we may greatly simplify the integration by using the reduced matrix elements. As shown in the previous section, this potential can be written as $V_4=\textbf{S}\cdot[\textbf{p}_{12}\times\textbf{r}_{12},f(r)]_+$, where $\textbf{S}=\textbf{s}_{1}+\textbf{s}_{2}$ and $f(r)$ is the spatial part of the potential with appropriate constants. One can write $\textbf{p}_{12}=-i\hbar\nabla_{r_{12}}$, and then $\nabla_{r_{12}} f(r_{12})=\textbf{e}_{12}\partial_{r_{12}} f(r_{12})$. We see, that when a gradient in the commutator operates on $f(r)$, we get $-i (\textbf{e}_{12}\times\textbf{e}_{12})\partial_{r_{12}} f(r_{12})=0$. We conclude, that $V_4$ may be written as
\begin{widetext}
\begin{eqnarray}
V_4=\textbf{S}\cdot[\textbf{p}_{12}\times\textbf{r}_{12},f(r_{12})]_+=2f(r_{12})\textbf{S}\cdot\left(\textbf{p}_{12}\times\textbf{r}_{12}\right)=-2f(r_{12})\textbf{S}\cdot\left(\textbf{r}_{12}\times\textbf{p}_{12}\right),
\end{eqnarray}
where we have used the fact that $(\textbf{p}_{12}\times\textbf{r}_{12})_i=\epsilon_{ijk} p^{j}_{12}r^{k}_{12}=\epsilon_{ijk} r^{k}_{12} p^{j}_{12}=-(\textbf{r}_{12}\times\textbf{p}_{12})_i$. The expectation value of this operator, needed to get $\Delta U_{ab,i}(m_0)$, can be obtained using reduced matrix elements. For state $|JMSL\rangle=|JM11\rangle$ we have
\begin{eqnarray}
&&\langle JM11|V_4|JM11\rangle=-\left\{ \begin{array}{lll}J&1&1\\ 1&1&1 \end{array}\right\} \langle S\|\textbf{S}\|S\rangle_{S=1} \langle L\|f(r_{12})\textbf{r}_{12} \times\textbf{p}_{12}\|L\rangle_{L=1}=\frac{J(J+1)-4}{2\sqrt{6}}\langle 1\|f(r_{12})\textbf{r}_{12}\times\textbf{p}_{12}\| 1\rangle,\nonumber\\
\label{eq:jm11reduced}
\end{eqnarray}
where we have introduced the 6j symbols \cite{Landau, Varshalovich} and used the fact that $\langle S\|\textbf{S}\|S\rangle=\sqrt{S(S+1)(2S+1)}$. Calculating the remaining reduced matrix element yields:
\begin{eqnarray}
\langle JM11|V_4|JM11\rangle=\frac{1}{2}[J(J+1)-4]\langle L| f(r_{12})\left( 1-D_{12}-D_{21} \right)| L\rangle_{L=1},
\end{eqnarray}
where
\begin{eqnarray}
D_{jk}=i r_j \sin \theta_j\left[\sin(\phi_j-\phi_k)\left(\sin\theta_k\frac{\partial}{\partial r_k}+\frac{\cos\theta_k}{r_k}\frac{\partial}{\partial\theta_k}\right)-\cos(\phi_j-\phi_k)\frac{1}{r_k\sin\theta_k}\frac{\partial}{\partial\phi_k}\right]
\end{eqnarray}
\end{widetext}
and $|L=1\rangle$ is a state represented by the first wave function in Eq. (\ref{eq:P1}) in the paper. This result was used to plot the Fig. \ref{fig:v4} of the paper.

\end{document}